\documentstyle[sprocl]{article}

\def\be{\begin{equation}}
\def\ee{\end{equation}}
\def\bea{\begin{eqnarray}}
\def\eea{\end{eqnarray}}

\begin{document}

\title{DARK MATTER, QUANTUM GRAVITY, \\
VACUUM ENERGY, AND LORENTZ INVARIANCE}

\author{ROLAND E. ALLEN}

\address{Department of Physics, Texas A\&M University \\
College Station, Texas 77843, USA \\
e-mail: allen@tamu.edu}


\maketitle\abstracts{We discuss the problems of dark matter, quantum
gravity, and vacuum energy
within the context of a theory for which Lorentz invariance is not
postulated, but instead emerges as a natural consequence in the physical
regimes where it has been tested.}

In earlier work~\cite{allen1,allen2}, we introduced a theory which implies
violation of Lorentz invariance for (i) fermions at extremely high energy
and (ii) fundamental scalar bosons which have not yet been observed. On the
other hand, the theory appears to be in agreement with even the most
sensitive experimental and observational tests of Lorentz invariance
that are currently available, since
many features of this symmetry are preserved, including rotational
invariance, CPT invariance, and the same velocity $c$ for all massless
particles.

\section{Dark Matter}

Let us begin with the dark matter problem. It appears that conventional
models of cold dark matter predict too much structure on small distance
scales~\cite{mondragon}. Since the dark matter almost certainly consists of
particles of a new kind, let us allow for the possibility that $v_{0}\neq 0$,
where $v_{0}$ is the limiting value of the particle velocity $v\left(
p\right) $ as the 3-momentum $\vec{p}$ goes to zero. Suppose that
the particle energy $\varepsilon $ is expanded as a Taylor series in the
magnitude $p$ of the 3-momentum:
\begin{equation}
\varepsilon =\varepsilon \left( p\right) =\varepsilon _{0}+pv_{0}+p^{2}/2
\widetilde{m}+\ldots .
\end{equation}
(For conventional nonrelativistic particles, $\widetilde{m}$ is
the particle mass, $\varepsilon _{0}$ is the rest mass energy, and $v_{0}=0$;
for particles with zero
rest mass, $v_{0}=c=1$ and $\varepsilon _{0}=\widetilde{m}^{-1}=0$.) The
particle velocity is then
\begin{equation}
v=d\varepsilon / dp=v_{0}+p/\widetilde{m}+\ldots
\end{equation}
and the kinetic energy is
\begin{equation}
T=\int v\,dp=\varepsilon \left( p\right) -\varepsilon _{0}=pv_{0}+p^{2}/2
\widetilde{m}+\ldots.
\end{equation}
The virial theorem implies that
\begin{equation}
\left\langle pv\right\rangle =\left\langle \vec{p}\cdot \vec{v}\right\rangle
=-\left\langle \vec{F}\cdot \vec{r}\right\rangle =\left\langle
r\,dV/dr\right\rangle =-\left\langle V\right\rangle
\end{equation}
where it has been assumed that $V=-GMm/r$ with $M$ constant. Since
(3) can also be written as
\begin{equation}
T=pv-\int p\,dv
\end{equation}
the binding energy -$E$ of a particle with 3-momentum $p$ is given by
the simple expression
\begin{equation}
-E=-\left\langle T+V \right\rangle =\left\langle \int
p\,dv\right\rangle =\left\langle p^{2}\right\rangle /2\widetilde{m}+\ldots
\approx \left\langle p^{2}\right\rangle /2\widetilde{m}.
\end{equation}
If $v_{0}=0$ (as for a conventional nonrelativistic particle), the momentum
is determined by
\begin{equation}
-\left\langle V\right\rangle =\left\langle pv\right\rangle =\left\langle
pv_{0}+p^{2}/\widetilde{m}+\ldots \right\rangle \approx
\left\langle p^{2}\right\rangle /\widetilde{m}\,\quad \mbox{or}\quad
\left\langle p^{2}\right\rangle \approx \left\langle GMm\widetilde{m}
/r\right\rangle
\end{equation}
and the energy has the familiar form
\begin{equation}
E\approx \left\langle V\right\rangle /2.
\end{equation}
On the other hand, if $v_{0}\neq 0$, the momentum is determined by
\begin{equation}
-\left\langle V\right\rangle =\left\langle pv\right\rangle =\left\langle
pv_{0}+p^{2}/\widetilde{m}+\ldots \right\rangle \approx
\,\left\langle p\right\rangle v_{0}\quad \mbox{or}\quad \left\langle
p\right\rangle \approx \left\langle GMm/v_{0}r\right\rangle
\end{equation}
and the binding energy is much smaller:
\begin{equation}
E\sim -\frac{1}{2\widetilde{m}v_{0}^{2}}\left\langle V\right\rangle ^{2}.
\end{equation}
The specific form of $\varepsilon \left( p\right) $ in the fundamental
theory of Refs. 1-3 yields
\begin{equation}
v_{0}=c\left[ 1+\left( \frac{2m}{\overline{m}}\right) ^{2}\right]
^{-1/2}\quad ,\quad \widetilde{m}v_{0}^{2}=mc^{2}\left( \frac{\overline{m}}
{2m}\right) ^{3}\frac{c}{v_{0}}.
\end{equation}
It is interesting, however, that a general model with $v_{0}\neq 0$ leads
to the
weaker binding (10), and thus to a weaker tendency to form both small-scale
structure and  cusps near the centers of galactic halos, apparently
in agreement with the observations.

\pagebreak
\section{Gravity and Gauge Fields}

Now let us turn to gravity and gauge fields, which have a radically new
interpretation in the present theory~\cite{allen1}:
The gravitational vierbein $e_{\alpha }^{\mu }$
is identified with the ``superfluid velocity'' $v_{\alpha }^{\mu }$ of a GUT
Higgs field which condenses in the very early universe.
The Euclidean action of this condensate initially has the form I(3.3):
\begin{equation}
S_{0}=\int d^{D}x\,\Psi _{s}^{\dagger }\left( T+\frac{1}{2}V
-\mu\right) \Psi_{s} \quad \mbox{with}\quad
V=b\Psi_{s}^{\dagger }\Psi_{s}.
\end{equation}
A local minimum in $S$ is given by
$\delta S=0$ for arbitrary variations $\delta \Psi _{s}$ and
$\delta \Psi _{s}^{\dagger }$. The arguments in
Section 3 of Ref. 1 then lead to a Bernoulli equation
$m\overline{v}^{2}/2+V+P=\mu$.
For an additional bosonic or fermionic field, the Euclidean action initially
has the form I(4.1),
\begin{equation}
S_{a}=\int d^{D}x\,\Psi_{a}^{\dagger }\left( T+V -\mu \right) \Psi_{a}
\end{equation}
if terms of order $\left( \Psi_{a}^{\dagger} \Psi_{a} \right)^{2}$ are
neglected. When $\Psi_{s}$ satisfies its equation of motion, 
the Bernoulli equation holds, and it can be used
in (13) to obtain the generalization of I(9.5) given in Ref. 2:
\begin{equation}
{\cal L}_{a}=\frac{1}{2}\widetilde{g}\left( -\bar{m}^{-1}\widetilde{g}^{\mu
\nu }D_{\mu }\psi _{a}^{\dagger }D_{\nu }\psi _{a}+i\psi _{a}^{\dagger}
e_{\alpha }^{\mu }\sigma ^{\alpha }D_{\mu }\psi _{a}\right) + h.c.
\end{equation}
Here $\psi_{a}$ is the field in a four-dimensional Lorentzian 
description, as observed in the 
frame of reference that is ``moving with the condensate". 
${\cal L}_{a}$ is actually an effective Lagrangian, which yields the 
same equation of motion for $\psi_{a}$ in a gravitational field 
as would be obtained from (13) 
(when $\Psi_{s}$ also satisfies its equation of motion). For fermions 
at low energy, this is exactly the same as in standard physics. The 
role of fermions and fundamental bosons as sources of gravity will be 
discussed elsewhere.

The Einstein field equations are also given by $\delta S=0,$ but this time
for variations in the metric tensor $g^{\mu \nu }$. In a Euclidean picture,
we search for a minimum in $S$ with respect to $g^{\mu \nu }$, while
remaining on the minimum with respect to 
$\Psi_{s}^{\dagger}$ that is represented by 
the equation of motion. (This is analogous to searching for the state 
of a particle with minimum energy $\varepsilon_{k}$, while requiring  
that $\psi_{k}$ always satisfy the Schr\"{o}dinger equation.) 
In a Lorentzian picture, we search for an extremum in the Lorentzian 
action $S_{L}$, while again requiring that $\Psi_{s}$ always satisfy 
its equation of motion.

In the present theory, the curvature of gravitational and gauge
fields can only result from topological defects, and it is these defects
which also give rise to the Einstein-Hilbert Lagrangian ${\cal L}
_{G}=\ell _{P}^{-2}g \, ^{(4)}R$ and the gauge-field Lagrangian
${\cal L}_{g}=-\left( 4g_{0}^{2}\right) ^{-1}gF_{\mu \nu }^{i}F_{\rho \sigma
}^{i}g^{\mu \rho }g^{\nu \sigma }$. (Here $g_{0}$ is the coupling constant
and $g=\left| \det \,e_{\mu }^{\alpha }\right|=
\left| \det \,g_{\mu \nu}\right|^{1/2}$.) In Ref. 1,
we considered defects with
point singularities. Suppose, however, that we assume (i) a short-distance
cutoff $a_{0}\sim \ell _{P}$ (which is implied by the microscopic treatment
of Ref. 2, with $\ell _{P}$ the Planck length) and (ii) a long-distance
cutoff $R_{0}$ (analogous to that in a superconductor) which results from
screening. With these assumptions, defects with line singularities --
i.e., vortex lines -- will have finite action per unit area in
four-dimensional Euclidean spacetime, in analogy with the finite energy per
unit length for a vortex line in a superconductor~\cite{goodstein}.
For example, consider a vortex line which has a length $\ell $
in 3-space and a duration $\Delta t$, so that the 4-dimensional
volume is $\sim \ell \Delta t \, \ell _{P}^{2}$ in units with
$c=1$. (These vortex lines will ordinarily be extended vortex rings, 
which can arise in the condensate, expand, and shrink back to 
zero over some finite period of time.) The contribution to the 
Euclidean action is then given by
\begin{equation}
\Delta S\propto \int d^{4}x\,n_{s}mv_{\theta }^{2}\propto \ell \Delta t
\left( n_{s}/m\right) \int d^{2}x\,r^{-2}\sim \left( \ell \Delta t /\ell
_{P}^{2} \right) \log \left( R_{0}/a_{0}\right) \sim \ell \Delta t /\ell
_{P}^{2}
\end{equation}
where $v_{\theta } = \left( mr\right) ^{-1}$ is the ``superfluid
velocity'' around the vortex. Also, the contribution to the square of
a gauge curvature is essentially given by
\begin{equation}
\ell \Delta t \int d^{2}x\left( \partial _{1}mv^{2}-\partial
_{2}mv^{1}\right) ^{2}\propto \ell \Delta t \int d^{2}x\,r^{-4}\,\propto
\ell \Delta t \left( a_{0}^{-2} - R_{0}^{-2} \right) \sim
\ell \Delta t /\ell_{P}^{2}.
\end{equation}
The contribution to the action is thus equal to the contribution to
(16) multiplied by a dimensionless constant of order unity.

The simplest example of a vortex line producing curvature and action is one
in which $mv_{i}^{\mu }$ is identified with $eA^{\mu }$, where $e$ is the
fundamental charge and $A^{\mu }$ is the electromagnetic vector potential
(in a convention which differs from that of Ref. 1 by the factor of $e$).
For example, with $B_{z}=\partial _{x}A_{y}-\partial
_{y}A_{x}$, a vortex line with a Planck-scale core makes a discrete
contribution to the magnetic flux $\Phi ,$ as well as to the action (15) and
to the quantity (16) which provides a measure of the action. In the present
picture, the magnetic flux through a surface is a time average of the
contributions from a large number of Planck-scale vortex lines.
The flux contributed by one vortex line is a flux quantum $\phi _{0}$:
$\Phi =\int_{S}dx\,dy\,B_{z}=\int_{C} A_{\theta } \,r\,d\theta
=\int_{C} e^{-1} r^{-1} \,r\,d\theta
=2\pi /e=2\pi \hbar c/e=\phi _{0}$, with $\hbar$ and
$c$ restored in the next to last expression.
Due to rapid fluctuations in the number and positions of these
topological defects, however, the contributions of individual defects cannot
be easily resolved, and the field appears to be continuous on length and time
scales which are large compared to $\ell _{P}$.

\pagebreak
This picture can be rather straightforwardly extended to the full
electromagnetic field, to nonabelian gauge fields with
$F_{\mu \nu }=\partial _{\mu }A_{\nu }-\partial _{\nu }A_{\mu }+\left[ A_{\mu
},\,A_{\nu }\right]$, to gauge fields in curved spacetime,
and to the gravitational field with~\cite{green}
\begin{equation}
\omega _{\mu }^{\alpha \beta }=\frac{1}{2}e^{\nu \alpha }\left( \partial
_{\mu }e_{\nu }^{\beta }-\partial _{\nu }e_{\mu }^{\beta }\right) -\frac{1}
{2}e^{\nu \beta }\left( \partial _{\mu }e_{\nu }^{\alpha }-\partial _{\nu
}e_{\mu }^{\alpha }\right) -\frac{1}{2}e^{\rho \alpha }e^{\sigma \beta
}\left( \partial _{\rho }e_{\sigma \gamma }-\partial _{\sigma }e_{\rho
\gamma }\right) e_{\mu }^{\gamma },
\end{equation}
$e_{\alpha }^{\mu} = v_{\alpha }^{\mu }$, $R_{\mu \nu }\,^{\alpha }
\,_{\beta }=\partial _{\mu }\omega \,_{\nu
}\,^{\alpha }\,_{\beta }-\partial _{\nu }\omega \,_{\mu }\,^{\alpha
}\,_{\beta }+\left[ \omega \,_{\mu }\,,\omega \,_{\nu }\right] ^{\alpha
}\,_{\beta }$, $R_{\mu \nu }\,^{\alpha }\,_{\beta }=
e_{\sigma}^{\alpha}e^{\tau}_{\beta}R_{\mu \nu }\,^{\sigma }\,_{\tau }$.
A detailed treatment will be given in a longer paper.

Notice that quantum gravity is finite in the present picture, since the
Einstein-Hilbert action is valid only on length scales that are large
compared to $\ell _{P}.$

\section{Vacuum Energy}

Finally, consider the vacuum stress-energy tensor
${\cal T}_{\mu \nu }^{vac}=
-\left( 2/g \right) \delta S_{vac} / \delta g^{\mu \nu}$. The equation of
motion for $\Psi _{s}$, and the Bernoulli equation below (12),
represent a local minimum in the Euclidean action $S$ ($\delta S=0$ for
arbitrary $\delta \Psi _{s}^{\dagger })$,
but they do not represent a minimum with
respect to variations in $g^{\mu \nu }$ ($\delta S=0$ as $g^{\mu \nu }$ is
varied, with $\Psi _{s}$ always required to satisfy its equation of motion).
There does not, in fact, appear to be any obvious reason why ${\cal T}_{\mu
\nu }^{vac}$ should be nearly zero within the simplest formulation of
the present theory. The vacuum energy problem therefore remains just as big
a mystery in the present theory as it is in standard field theory and in
superstring/M theory.

\section*{Acknowledgement}

This work was supported by the Robert A. Welch Foundation.

\end{document}